\documentclass[journal=jctcce,manuscript=article]{achemso}

\newcommand{\onlinecite}[1]{\hspace{-1 ex} \nocite{#1}\citenum{#1}} 
\DeclareUnicodeCharacter{308}{{\"u}}

\usepackage{graphics,graphicx,amsfonts,amsmath,amsbsy,amssymb,color}
\usepackage{verbatim}  
\usepackage{psfrag}  
\usepackage{tabularx}
\usepackage{xmpmulti}
\usepackage{hyperref}
\usepackage{marginnote}
\usepackage{subfigure}
\usepackage{paracol}
\usepackage{xparse}
\usepackage{layouts}
\usepackage{afterpage}
\usepackage{braket}
\usepackage{paralist}
\usepackage{bm}
\usepackage{url}
\usepackage[normalem]{ulem}
\usepackage{color}
\usepackage[colorinlistoftodos]{todonotes}
\usepackage[american]{babel}

\newcommand{\reffig}[1]{{Fig.~\ref{#1}}}

\newcommand{\fcut}{\ensuremath{f_\mathrm{cut}}}

\hypersetup{hyperindex,breaklinks,colorlinks=true,linkcolor=blue,citecolor=blue,urlcolor=blue,filecolor=blue}

\title{Power laws used to extrapolate the coupled cluster correlation energy to the thermodynamic limit}
\author{Tina~N~Mihm}%
\affiliation{Department of Chemistry, University of Iowa}%
\author{Bingdi~Yang}%
\affiliation{Department of Chemistry, University of Iowa}
\author{James~J.~Shepherd}%
\affiliation{Department of Chemistry, University of Iowa}%
\email{james-shepherd@uiowa.edu}
\date{\today}

\begin{document}

\begin{abstract}
Recent calculations using coupled cluster on solids have raised discussion of using a $N^{-1/3}$ power law to fit the correlation energy when extrapolating to the thermodynamic limit, an approach which differs from the more commonly used $N^{-1}$ power law which is (for example) often used by quantum Monte Carlo methods. 
In this paper, we present one way to reconcile these viewpoints.  
Coupled cluster doubles calculations were performed on uniform electron gases reaching system sizes of $922$ electrons for an extremely wide range of densities ($0.1<r_s<100.0$) to study how the correlation energy approaches the thermodynamic limit. %
The data were corrected for basis set incompleteness error and use a selected twist angle approach to mitigate finite size error from shell filling effects. 
Analyzing these data, we initially find that a power law of $N^{-1/3}$ appears to fit the data better than a $N^{-1}$ power law in the large system size limit.
However, we provide an analysis of the transition structure factor showing that $N^{-1}$ still applies to large system sizes and that the apparent $N^{-1/3}$ power law occurs only at low $N$.

\end{abstract}
\maketitle

\section{Introduction}

High-accuracy electronic structure methods such as coupled cluster theory are being advanced as tools for materials design~\cite{manby_extension_2006, stoll_incremental_2009, voloshina_correlation_2007, booth_towards_2013, gruneis_natural_2011, irmler_duality_2019, gruber_applying_2018, irmler_particle-particle_2019, hummel_low_2017, zhang_coupled_2019, gruneis_efficient_2015,lewis_ab_2019,mcclain_gaussian-based_2017, motta_hamiltonian_2019, pulkin_first-principles_2020, sun_gaussian_2017}.
The value of coupled cluster theory is its ability to determine correlation energy, $E_\mathrm{total}-E_\mathrm{HF}$ where $E_\mathrm{HF}$ is the Hartree--Fock energy, in an extremely versatile manner. 
Despite considerable advances, a remaining hurdle to the widespread effectiveness of these methods is their computational scaling. 
For meaningful insights into solid-state systems, calculations must be performed in or extrapolated to the thermodynamic limit (TDL)---the limit of an infinite number of atoms or k-points---often at considerable cost.
Coupled cluster methods tend to scale with system size or k-points with a high polynomial cost; this is magnified by the need to use large basis sets for highly-accurate results.
There is, therefore, a critical need to understand how the coupled cluster correlation energy approaches the thermodynamic limit with increasing system size. 

A common method to reach the TDL via extrapolation is to run several large $N$ calculations, and then to fit the energies per particle as a function of $N$ to $E_N= E_\mathrm{TDL}+AN^{-\alpha}$, where $\alpha$ is some positive constant commonly considered to be consistent between 3D solids. %
The energy can change significantly as a function of N and, in general, exhibits complex polynomial behavior. 
Simple, single-term power law expansions become visible only when the system size is large enough.
If $\alpha$ is known from first principles, the situation can be improved, as (1) the extrapolation is more likely to be correct, and (2) the extent to which the data fit the power law indicates to what extent the extrapolation is likely in error.

In general, it has been assumed that the coupled cluster correlation energy follows the same trends on approach to the thermodynamic limit of other methods. 
The $N^{-1}$ dependence of the energy is routinely used for TDL extrapolations in solid-state for some coupled cluster theory calculations~\cite{liao_communication:_2016, liao_comparative_2019, marsman_second-order_2009, gruber_ab_2018,booth_towards_2013}, where the physical argument which relates this power law to coupled cluster involves how long-range van der Waals interactions behave in extended systems~\cite{gruber_applying_2018} and analysis of the transition structure factor~\cite{liao_communication:_2016}.
This power law is widely used to extrapolate quantum Monte Carlo calculations~\cite{fraser_finite-size_1996, williamson_elimination_1997, kent_finite-size_1999, lin_twist-averaged_2001, chiesa_finite-size_2006, gaudoin_quantum_2007, drummond_finite-size_2008, kwee_finite-size_2008,dornheim_ab_2016, ceperley_ground_1987, kwon_effects_1998, foulkes_quantum_2001, gurtubay_benchmark_2010, filinov_fermionic_2015, shepherd_many-body_2013, ruggeri_correlation_2018, holzmann_theory_2016, booth_towards_2013, ceperley_ground_1978}.
Recently, some coupled cluster calculations have been made where authors observe a different power law of $N^{-1/3}$;~\cite{mcclain_gaussian-based_2017, motta_hamiltonian_2019, pulkin_first-principles_2020, sun_gaussian_2017} in these calculations periodic Gaussian basis sets are used.
These studies offer a physical explanation which suggests that the slower power law which occurs in the Hartree--Fock energy when a Madelung constant is not used also affects the correlation energy due how the Hartree--Fock eigenvalues behave~\cite{mcclain_gaussian-based_2017}. 

An $N^{-1/3}$ power-law would imply a significantly slower convergence to the TDL than $N^{-1}$ and so we thought it worthy of investigation.
To investigate, we collected coupled cluster doubles (CCD) data for the uniform electron gas (UEG) at eight $r_s$, ranging from $0.1$ to $100.0$, for system sizes up to $N=922$.
We use complete basis set limit corrections~\cite{shepherd_communication:_2016} and a selected twist angle approach to reduce fluctuations in the energy from shell filling effects.~\cite{mihm_optimized_2019}
Initially, we found that a $N^{-1/3}$ power law provided a better numerical fit to the data than an $N^{-1}$ power law.
Through analyzing the transition structure factor, we found that the apparent $N^{-1/3}$ convergence rate occurs due to simulating a supercell that is too small and does not accurately capture the convergence of the structure factor to the TDL (i.e. $N\rightarrow \infty$). 
This provides an explanation for how observed trends in the energy of $N^{-1/3}$ (Ref.~\onlinecite{mcclain_gaussian-based_2017}) may become $N^{-1}$ at higher $N$.
We then discuss the implications of our work.%

\section{Methods}

\subsection{Coupled cluster doubles on the uniform electron gas}

We follow methods and procedures outlined more thoroughly elsewhere~\cite{shepherd_range-separated_2014, shepherd_coupled_2014,shepherd_many-body_2013}; however, for the benefit of the reader, we will highlight some key methodological details here.
Coupled cluster theory uses an exponential ansatz for the wavefunction: $\Psi = e^T \Psi_\mathrm{HF}$. 
The energy is found by projection: $E = \langle \Psi_\mathrm{HF} | H | e^T  \Psi_\mathrm{HF} \rangle$ with the aim of finding the correlation energy.
To obtain a method with manageable computational cost, the $T$ operator is truncated at a certain excitation level. 
Here, we truncate at the doubles level and this method is called coupled cluster doubles (CCD). 

This formulation yields self-consistent equations for the individual elements $t_{ijab}$ of the operator $T$, and an expression for the correlation energy, $E_\mathrm{corr} =$ $E_\mathrm{total}-E_\mathrm{HF}$, as follows:
\begin{equation}
E_\mathrm{corr}=\frac{1}{4} \sum_{ijab} t_{ijab} \bar{v}_{ijab}
\end{equation}
where $\bar{v}_{ijab}$ are antisymmetrized four-index electron repulsion integrals. Here, $i$ and $j$ refer to occupied orbitals, and $a$ and $b$ refer to virtual orbitals over some finite basis set. 
This energy expression is related to the transition structure factor recently coined by Gr\"uneis and coworkers, $S_G=\sum_{ijab}^\prime (2t_{ijab}-t_{jiab})$, where the prime indicates that the sum is only taken over excitations that are related by the momentum transfer $G$.~\cite{irmler_duality_2019,liao_communication:_2016,gruber_applying_2018}

In the electron gas, all orbitals are plane waves: $\phi_j \propto \exp({\overline{i} {\bf k}_j \cdot {\bf r}}$), where ${\bf k}_j$ is a vector of the momentum quantum numbers in 3D, ${\bf r}$ is the electron coordinate, and $\overline{i} = \sqrt{-1}$.
A simple cubic three-dimensional box is used with electron numbers which correspond to closed-shell configurations at the $\Gamma$-point.
We use the Ewald interaction (per convention) for calculations with periodic boundary conditions.
This gives rise to matrix elements $v_{ijab}$ (electron repulsion integrals) that take the familiar form $1/G^2$, where $G$ is the magnitude of the momentum transfer during the {$i$,$j$} to {$a$,$b$} excitation, provided the excitation is allowed by momentum conservation (i.e., ${\bf k}_a-{\bf k}_i={\bf k}_j-{\bf k}_b={\bf G}$). 
We explicitly calculate and include a Madelung term, $v_M$ in our calculations~\cite{fraser_finite-size_1996}. 
For example, at a Wigner-Seitz radius $r_s=1.0$ and $N=14$, $v_M=0.7303$~\footnote{This term represents the exchange energy of an electron interacting with its own periodic image with a neutralizing background. This value is provided for other practitioners to reproduce our numbers; $v_M$ at all densities and box lengths can be calculated by re-scaling this value}.
A finite basis set for coupled cluster calculations is defined by the $M$ orbitals that lie inside a kinetic energy cutoff $E_{cut, m} = \frac{1}{2}k_\mathrm{cut}^2$. 

The Hartree--Fock eigenvalues for the occupied and virtual orbitals, following our previous work~\cite{shepherd_coupled_2014}, are: 
\begin{align}
\epsilon_p = \left\{
\begin{array}{ll}
\frac{1}{2} {\bf k}_p^2 - \sum_{\substack{j \in \textrm{occ}\\p \neq j}}  v_{pj} - v_M , &\quad p \in \textrm{occ} \\
\frac{1}{2} {\bf k}_p^2 - \sum_{\substack{j \in \textrm{occ}}}  v_{pj} , &\quad p \in \textrm{virt}.
\end{array}
\right. %
\end{align}
In these equations $\frac{1}{2} {\bf k}_p^2$ is the kinetic energy of an electron in orbital $p$ and $v_{pj}$ is the exchange energy between orbitals $p$ and $j$.
We note that the occupied orbitals are each lowered in energy by $v_M$, the Madelung constant.
This represents the exchange interaction of an electron with itself. In the thermodynamic limit, $v_M\rightarrow 0$. 

These calculations were performed with a locally modified version of a github repository used in previous work http://github.com/jamesjshepherd/uegccd ~\cite{shepherd_range-separated_2014,shepherd_coupled_2014}.
Also, Hartree atomic units are used throughout this paper.

\subsection{Connectivity twist averaging}

Twist averaging is employed to reduce the fluctuations in the energy with respect to system size on converging to the thermodynamic limit~\cite{lin_twist-averaged_2001, gruber_ab_2018, gruber_applying_2018, drummond_finite-size_2008, liao_communication:_2016, maschio_fast_2007, zong_spin_2002, pierleoni_coupled_2004, mostaani_quantum_2015, azadi_efficient_2019}.
In general, this amounts to offsetting the grid of momenta by applying a twist angle, ${\bf k}_s$, to each orbital such that:
$\phi_j \propto \exp(\overline{i} ({\bf k}_j - {\bf k}_s)\cdot {\bf r})$,
and then averaging the correlation energy over all twists angles such that:
\begin{equation}
\langle E_\mathrm{corr} \rangle_{{\bf k}_s}=\frac{1}{N_s}\sum_{t=1}^{N_s}E_\mathrm{corr}({\bf k}_{s_t})
\end{equation}
where $N_s$ is the number of twist angles used in the twist averaging. 

Instead of calculating this costly sum explicitly, we instead use the connectivity twist averaging scheme. 
This method estimates a single twist angle for each calculation.
This is chosen by considering the available momentum transfer vectors between the occupied space and the virtual space to second order (which we termed the connectivity).
A twist angle is chosen by considering which twist best matches the average connectivity. 
Overall, this allows us to compute a reliable approximation to the twist-averaged energy using only one calculation~\cite{mihm_optimized_2019}. 
The connectivity twist averaging scheme reduces the computational cost by a factor of approximately $N_s$, which is vital for our work here.

Our recent work has shown that cTA can be used in calculations with methods that describe correlation more completely, such as QMC.\cite{mihm_accelerating_2021} 
Furthermore, when we ran cTA on different system sizes and basis sets, the calculated twist angle varied slightly. It may be possible to transfer the twist angle between different systems, but we do not investigate that here.

\subsection{Removing basis set error} 

Running calculations in a finite basis set incurs an error referred to as basis set incompleteness error (BSIE). Previous work has shown that, for a three-dimensional solid, this error decays as $1/M$ as $M\rightarrow \infty$~\cite{shepherd_communication:_2016, shepherd_range-separated_2014, shepherd_investigation_2012, shepherd_convergence_2012}. 
In general, to study convergence to the thermodynamic limit, we must first eliminate BSIE by ensuring that energies are converged with respect to basis set size for each electron number N; then, we must find the way that these energies asymptote as $N \rightarrow \infty$.
Performing this extrapolation in both $M$ and $N$ by brute force is impractical.
Instead, we derive a term to correct for basis set incompleteness error that we may apply before analyzing convergence to the thermodynamic limit. 

First, we judiciously choose a single value of $M$ for each particle number $N$~\cite{shepherd_many-body_2013} which allows for an approach to the thermodynamic limit using a finite basis set cutoff.  To do so, we ensure that a ``cutoff factor," $f_\mathrm{cut}$, is consistent for each calculation.  We define this cutoff factor as the ratio between the maximum kinetic energy of the virtual orbitals and the maximum kinetic energy of the occupied orbitals i.e. $f_\mathrm{cut}= E_{\mathrm{cut},M}/E_{\mathrm{cut},N}$. 

\begin{figure}

\includegraphics[width=0.5\textwidth,height=\textheight,keepaspectratio]{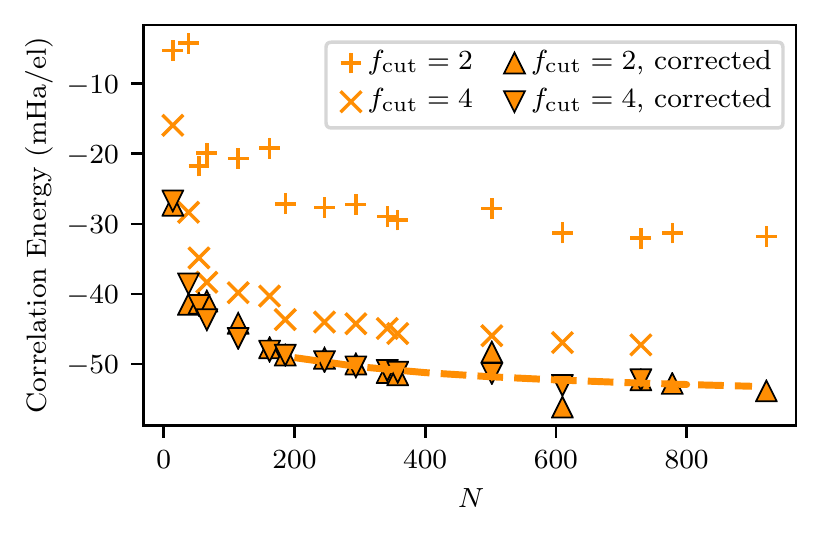}

\caption{Correlation energy per particle data for $r_s = 1.0$ at a wide range of $N$ ($14$ to $922$) are shown with a correction for the basis set error for the two basis sets (\fcut $= 2$ and $4$). After the correction, the data sets can be seen to lie on the same trend line with $N$. A fit line has been added to help guide the eye.}
\label{fig:Basis_set_cor}
\end{figure}

Second, we correct each calculation with a basis set correction factor noting that extrapolations in $M$ and $N$ tend to be independent and commute.~ \cite{shepherd_communication:_2016}
When the energy is written as a function of the the number of basis functions per electron ($m=M/N$), identical convergences to the complete basis set limit are seen regardless of the value of N (i.e., $E(N,m)\sim B(1/m) + E_\mathrm{CBS}(N)$). %
It is a logical extension of this argument that the complete basis set energy for any one electron number can be used to correct all electron numbers to the complete basis set limit. 
In this work, we used the complete basis set energy at an electron number of $N=186$ to correct all other data at the same $r_s$ to the complete basis set limit.
The $N=186$ system size was chosen because it was the largest system size whose CBS limit could be found using calculations on a single node in one day.
The following expression was used for the correction: 
\begin{equation}
\label{correction}
E^\prime(N,m)=E(N,m)-E(186,m)+E_\mathrm{CBS}(186).
\end{equation}
Care needs to be taken since $E(186,m)$ does not exist for all values of $m$. %
To get around this, we assume that we can linearly interpolate between two nearby calculations to find a specific value of $E(N,m)$:
\begin{align}
E(N,m)  &= E(N,m_2) \notag \\
&+ \frac{1/m - 1/m_2}{1/m_1 - 1/m_2}\left(E(N,m_1) - E(N,m_2)\right).
\end{align}
Data supporting this approach can be found in \reffig{fig:Basis_set_cor} and our prior work~\cite{shepherd_communication:_2016}.

\section{Results}

\begin{figure}

\subfigure[\mbox{}]{%
\includegraphics[width=0.5\textwidth,height=\textheight,keepaspectratio]{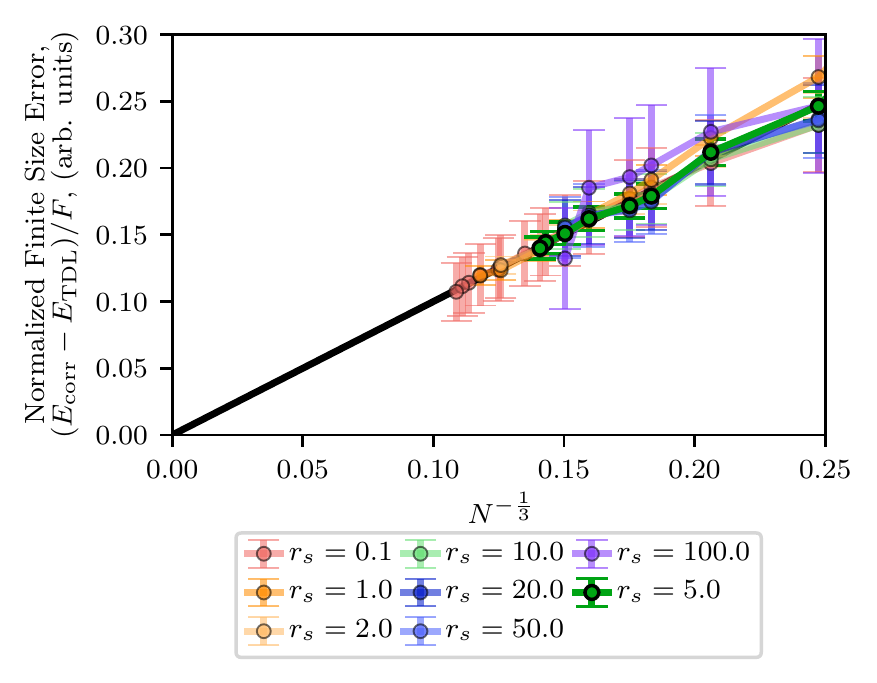}
\label{subfig:N-1_3-All_Rs}
}

\subfigure[\mbox{}]{%
\includegraphics[width=0.5\textwidth,height=\textheight,keepaspectratio]{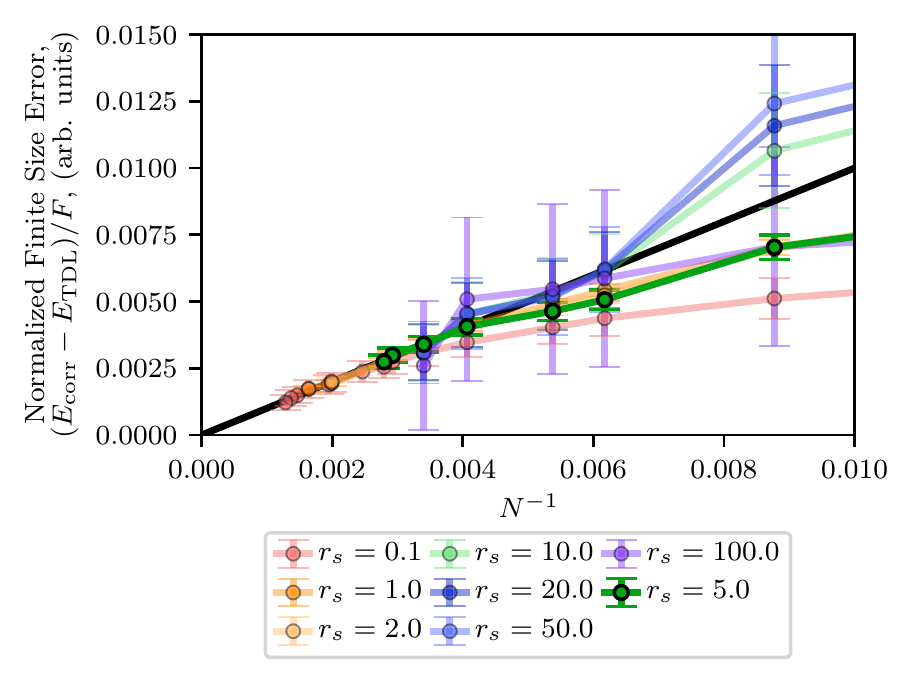}
\label{subfig:N-1-All_Rs}
}
\caption{Correlation energy data are fit to $E_\mathrm{corr} = E_\mathrm{TDL}+F N^{-\alpha}$ for (a) $\alpha=1/3$ and (b) $\alpha=1$. Here, the correlation energies are per particle. The energies are plotted as $(E_\mathrm{corr} - E_\mathrm{TDL})/F$ against $N^{-\alpha}$ as this highlights the rate at which energies converge to the extrapolation power law. 
Due to re-scaling the energy units are lost and we describe these as arbitrary units.
The fit was taken over the largest-$N$ $6$ to $8$ data points (depending on the $r_s$) to obtain $F$ and $E_\mathrm{TDL}$. The fit procedure minimizes the sum of the squared residuals to the fit function. Error bars are then calculated from the variance of the parameter estimates and propagation of error.
The black line indicates $(E_\mathrm{corr} - E_\mathrm{TDL})/F = N^{-\alpha}$. The smallest system visible on the graph is $N=114$.
Different $r_s$ values converge at different rates to both power laws.
}
\label{fig:E_slope_vs_N}
\end{figure}

We collected data for eight $r_s$ values, each 
with a set range of $N$ that were run with an $\fcut = 4$ to obtain convergence to the TDL. 
For $r_s$ of $0.1$, $1.0$, $2.0$, and $5.0$, we used $N$ ranges from $N=14$ up to $N=778$, $N=730$, $N=502$, $N=358$ respectively. 
For the remaining $r_s$ between $10.0$ to $100.0$, a set $N$ range of $N=14$ to $294$ was used. 
For the basis set corrections, most $r_s$ were run over $N=186$ and an $M$ range of $M=294$ to $7774$. The exceptions to this were $r_s = 10.0,~20.0$ and $100.0$. Here, the basis set ranges only went up to $M=8338,~5938,~5034$ and $7390$ respectively. 

\begin{figure}
	\includegraphics[width=0.5\textwidth,height=\textheight,keepaspectratio]{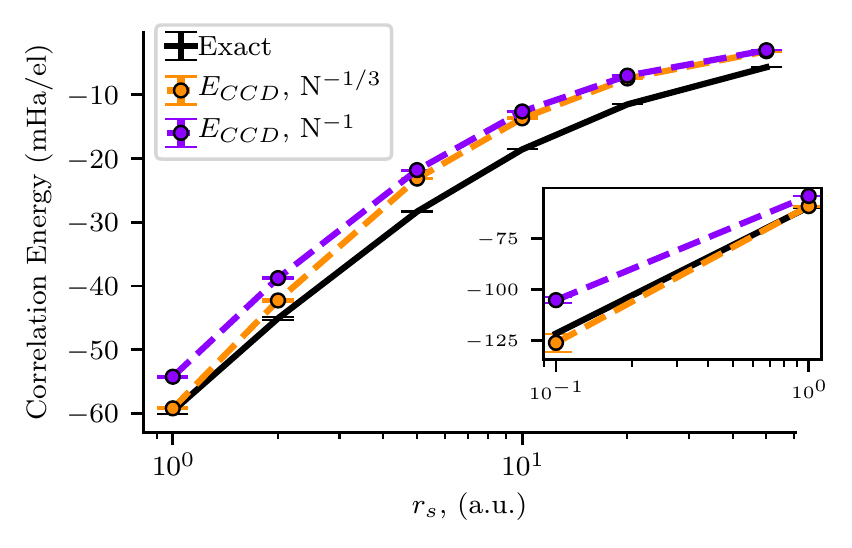}
	\caption{The thermodynamic limit energies for the two power laws are shown in comparison with exact values for a range of $r_s$ from $0.1$ to $50.0$. The smaller $r_s$ data have been graphed as an inset to better show the comparison between the energies for all $r_s$. The exact energy values were calculated using the Ceperley and Alder results for $r_s > 1.0$ and Gell-Mann and Brueckner results for $r_s < 1.0$ as provided by Perdew and Zunger \cite{perdew_self-interaction_1981, ceperley_ground_1980, gell-mann_correlation_1957, onsager_integrals_1966}. Errors for the exact energies were taken from Ceperley and Alder.~ \cite{ceperley_ground_1980}
	}
	\label{fig:TDLComp}
\end{figure}

\begin{figure}%
	\includegraphics[width=0.5\textwidth,height=\textheight,keepaspectratio]{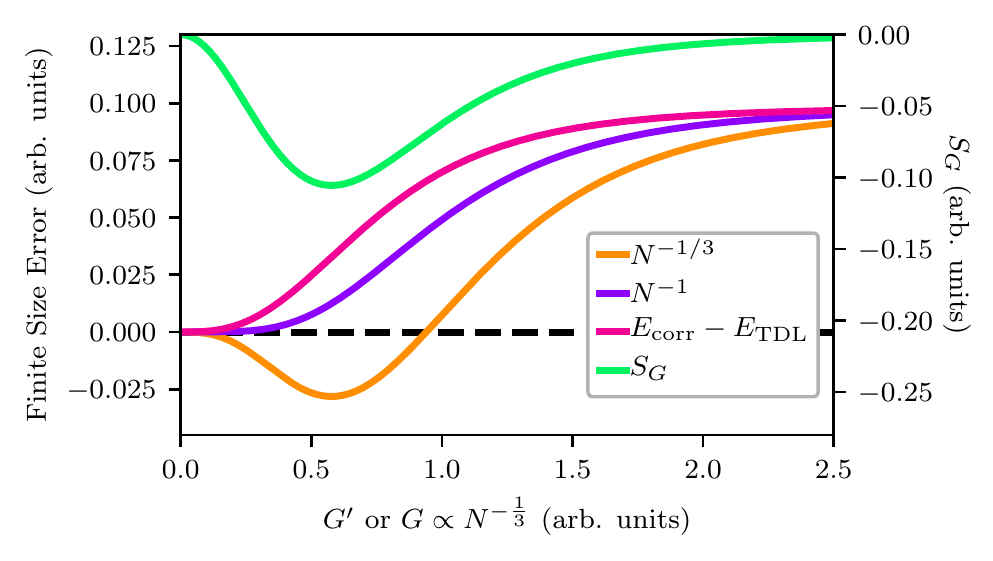}
	\caption{An analytical model for the transition structure factor is used to derive finite-size effects remaining after extrapolation to the thermodynamic limit (TDL), in order to compare the effectiveness of $N^{-1/3}$ and $N^{-1}$ extrapolations. The minimum in the transition structure factor (shown with the green line), serves as a cross-over point that indicates which power law gives a better TDL extrapolation. Symbolic manipulations were performed in Mathematica \cite{wolfram_research_inc_mathematica_2019}. As we were not attempting to fit these curves to real data, all units here are arbitrary units.}
	\label{fig:SFa}
\end{figure}

\begin{figure}%

	\includegraphics[width=0.6\textwidth,height=\textheight,keepaspectratio]{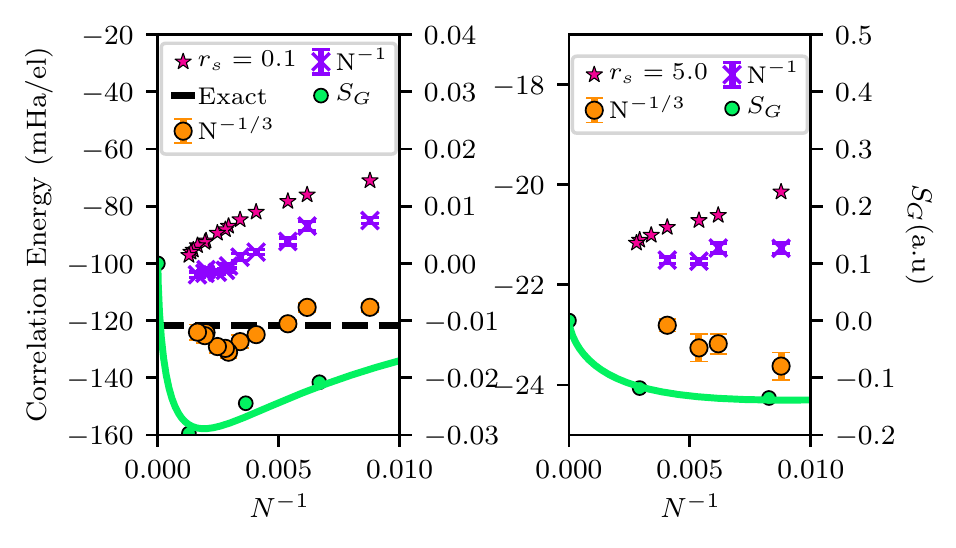}
	\label{subfig:SFb_rs5}
	\caption{CCD data for $r_s=0.1$ and $r_s=5.0$ are extrapolated to the thermodynamic limit. For each $N$, data in the extrapolated curve (labelled with the power-law) is derived from considering the previous 5 points. Error bars are calculated from the variance of the parameter estimates and propagation of error. Colors match \reffig{fig:SFa}. The continuous representation for the transition structure factor have been fit using the analytical formula presented in the text for both $r_s$.}
	\label{fig:SFb}
\end{figure}

The data we collected was used to investigate whether an $N^{-1}$ power law or an $N^{-1/3}$ power law was best for extrapolating to the TDL. 
The former is the more commonly-used power law and the latter is a recent observation by other coupled cluster users~\cite{mcclain_gaussian-based_2017}.
To do this, we constructed graphs where the independent axis is the expected power-law dependence of the energy. 
Such graphs are shown in \reffig{subfig:N-1_3-All_Rs} and \reffig{subfig:N-1-All_Rs}, where plots of the correlation energy for $r_s=1.0-100.0$ are shown. 
When we compare these two figures, we can see that the $N^{-1/3}$ power law fits the data in a way that is \emph{apparently} a better fit over $N^{-1}$.
For example, the onset to a linear regime is sooner and more straight for the $N^{-1/3}$ power law fits across all densities, while the $N^{-1}$ power law fits only show faster and improved convergence with increasing density. 

The total correlation energies after extrapolation are shown in \reffig{fig:TDLComp}. 
It can be seen that the two power laws lead to similar TDL energies at densities of $r_s>5$ i.e. for densities which correspond to real materials. 
The similarity of the result depends, though, on having reached a system size large enough where $N^{-1}$ produces a reasonable extrapolation.
At higher densities, the difference between the two power laws becomes more pronounced. 
At ultra high densities, as coupled cluster theory approaches the exact limit (following the random phase approximation) it even appears as though $N^{-1/3}$ is even required to reach the exact TDL. 

From this data alone, we would expect that $N^{-1/3}$ would be the correct power law and this conclusion would be at odds with the idea that there is a prevailing $N^{-1}$ power-law in the TDL. 
The $N^{-1/3}$ power law would, after all, be slower than $N^{-1}$ and would be seen in the limit of large $N$ if both were present. 
Absent other data, we might be tempted to conclude this is some unfortunate feature of coupled cluster theory.

In order to explain this seeming contradiction, we analyzed a property of the coupled cluster wavefunction called the transition structure factor, which was recently introduced in the literature~\cite{liao_communication:_2016, irmler_duality_2019, gruber_applying_2018}. 
The transition structure factor $S_G$ is defined by re-writing the correlation energy expression as: $E_\mathrm{corr}=\frac{1}{4}\sum_G t_{ijab} v_{ijab}=\sum_G S_G V_G$, where $V_G$ is the electron repulsion integral; both $S_G$ and $V_G$ are Fourier components labelled with a momentum $G$.
In the $S_{G\rightarrow0}=0$ limit, $S_G\sim G^2$ can be shown to give rise to the $N^{-1}$ relationship derived in the literature~\cite{mattuck_guide_1992,holzmann_theory_2016,holzmann_finite-size_2011}.

To begin, we use the following functional form to fit the transition structure factor from our coupled cluster doubles calculations:
\begin{equation}
    S_G \propto \frac{1}{(G^2+\lambda^2)^4}G^2.
\end{equation}
Our inspiration for this fit is that it comes from a screened potential, $\tilde{v}=e^{-\lambda r}$,~\cite{gruneis_explicitly_2013} and includes a $G^2$ term from the sums over the occupied manifold at small  $G$~\cite{mattuck_guide_1992}.
This functional form fits the numerically-determined structure factor for the UEG. %
If the smallest $G$ present in the calculation is given by $G^\prime$, then
\begin{equation}
F(G^\prime)=\int_{0}^{G^\prime} S_G V_G \, G^2 dG
\end{equation}
(where the factor $G^2$ comes from the G-space volume element in 3D) gives the finite-size error in the correlation energy. 
Since $G^\prime=\frac{2\pi}{L}
\propto N^{-1/3}$, the $G^\prime$ derivative of $F(G^\prime)$ gives the slope of the convergence into the TDL if a $N^{-1/3}$ power law is assumed. 
The remaining FSE after extrapolation is:
\begin{equation}
F(G^\prime)-G^\prime \frac{d}{d G^\prime} F(G^\prime).
\end{equation}
An analogous expression can be evaluated for an extrapolation based on $G^3\propto N^{-1}$.

Figure~\ref{fig:SFa} shows the results of this analysis, taking $\lambda=1$ (and ignoring energy prefactors for simplicity, yielding arbitrary units). 
Two regimes can be seen on either side of the minimum in $S(G^\prime)$. 
For $G^\prime$ values to the right of the minimum (i.e., when the system size is small), the $N^{-1/3}$ extrapolation falls with increasing $N$, and reaches the thermodynamic limit more quickly than the $N^{-1}$ extrapolation. However, as $G^\prime$ decreases and the system size gets larger, the $N^{-1/3}$ extrapolation eventually overshoots the true thermodynamic limit. 
For $G^\prime$ to the left of the minimum (i.e., when the system size is large) the $N^{-1/3}$ extrapolation rises with increasing $N$. 
Although $N^{-1}$ is analytically the correct extrapolation, at large $G^\prime$, it gives larger errors than the $N^{-1/3}$ extrapolation. 

The interpretation of this analysis is that the $N^{-1/3}$ is an apparent power law rather than the true asymptote which can be seen as a result of insufficiently sampling the low $G$ part of the transition structure factor where large enough system sizes will eventually give rise to $N^{-1}$ behavior. 

These analytical results show a remarkable agreement with our real data, shown in \reffig{fig:SFb}. 
In both $r_s=0.1$ and $r_s=5.0$, the $N^{-1/3}$-extrapolated result first falls, and then rises again.
For $r_s=0.1$, the exact result is known from the random phase approximation~\cite{gell-mann_correlation_1957}, and coupled cluster can be expected to converge to that result in the TDL~\cite{shepherd_coupled_2014,scuseria_ground_2008}. %
This provides a demonstration that the correlation energy appears to lie between the $N^{-1/3}$ and $N^{-1}$ extrapolations, again matching our analytical model above.

\section{Discussion}

To discuss how these results apply to real systems, we can find the $G$ (and thus the system size, $N$) corresponding to the minimum in the transition structure factor for each density.
For metallic densities between $r_s=5.0$ and $r_s=20.0$, the minimum was found at $N=100$, and the onset of the best fits to the $N^{-1}$ power law required calculations using $150<N<250$. 
Applying these numbers to real systems means that with a lithium unit cell, which is a simple metal and closely modelled by a uniform electron gas, we would expect to need $k$-point meshes finer than $5 \times 5 \times 5$ for a straight-forward $N^{-1}$ extrapolation. 
For smaller systems, the extrapolation from a $N^{-1/3}$ power law is likely to act as a lower bound and that of $N^{-1}$ an upper bound.

One limitation of our study is that the minimum in the transition structure factor (and thus our findings around these extrapolations) is affected by basis set incompleteness error. As the basis set is made more complete, the exact position of the minimum in $G$ will vary. 
As for the energies, though we have been careful with our basis set extrapolations, we cannot rule out various power-laws (tending to have fractional powers of $N$) coming from residual basis set incompleteness error or the extrapolation in $M$ not commuting with the extrapolation in $N$. 

Another limitation of this study is that we do not discuss correction terms which are a popular way to address some of the difficulties extrapolations pose.
There are many schemes for how to correct correlation energies and many are quite sophisticated~\cite{ drummond_finite-size_2008, kwee_finite-size_2008,  kwon_effects_1998, gurtubay_benchmark_2010, ruggeri_correlation_2018}. %
Our justification for not focusing on this was that our goal was to study the approach of the correlation energy to the TDL without such corrections to better follow the recent discussion in the literature.
It is also the case that we do not yet have a clear route towards including correction terms in the transition structure factor. 
We agree with the authors of a recent paper~\cite{ruggeri_correlation_2018} who indicate that the appearance of an $N^{-1/3}$ power law can also arise from there being a $N^{-2/3}$ term opposite in sign to $N^{-1}$. %
More generally, a limitation of our analysis is that it exclusively focuses on the interaction (through consideration of the transition structure factor) and does not account for the fractional power-laws from integration over grids.
Despite these limitations, our paper nonetheless provides an alternative and valuable perspective.%

It is also interesting to consider how applicable our findings here are beyond coupled cluster theory. 
The coupled cluster amplitudes, $t_{ijab}$, are related to the true wavefunction: When coupled cluster theory is exact (such as at high densities), these amplitudes are the coefficients in the intermediate-normalized full configuration interaction (FCI) wavefunction, i.e. $t_{ijab}\rightarrow c_{ijab}$ where $c_{ijab}=\langle \Psi_0 | \hat{a}^\dagger_a \hat{a}^\dagger_b \hat{a}_i \hat{a}_j   |  \Psi_{\mathrm{HF}}\rangle$; $\hat{a}$ are creation and annihilation operators and $\Psi_{\mathrm{HF}}$ is the Hartree--Fock (HF) wavefunction. Thus, the transition structure factor is a property (albeit non-observable) of the ground-state in any post-HF method, suggesting that our findings should be broadly transferable.
Though it is beyond the scope of this study, we would also like to understand whether the physical limit of $S_{G\rightarrow0}=0$ is only applicable in the infinite system result and whether there is a Madelung-like term in $S_G$. This would contribute through the Madelung term in the potential $V_M$ with both vanishing in the TDL. Regardless of the scaling of the power law, it is possible that a Madelung-like approach can be successful in remedying the error in the correlation energy where an $S_G$ term related to $G=0$ is included in truncations of $S_G$.

\section{Conclusions}

In summary, we have performed coupled cluster calculations on the uniform electron gas at a wide range of system sizes and densities. 
We then tested the effectiveness of two ways to extrapolate energies into the thermodynamic limit.
This study was motivated by a recent dilemma that has arisen in the literature around whether an $N^{-1/3}$ or an $N^{-1}$ power law matches convergence of the correlation energy to the thermodynamic limit.
The complete-basis-set-corrected twist-averaged correlation energies we calculated initially appeared to fit $N^{-1/3}$ better than $N^{-1}$ in their convergence to the thermodynamic limit. 
Going beyond this observation, we developed a heuristic analysis of the transition structure factor that explains the observation of an apparent $N^{-1/3}$ power law (as in Ref~\onlinecite{mcclain_gaussian-based_2017}) as resulting from using too small an electron number to resolve the minimum in the transition structure factor (in other words, insufficiently small $G$ to show the appropriate screening in the interaction). 
The power law reverts to a $N^{-1}$ behavior at large $N$.

We believe this conclusively explains recent observations related to $N^{-1/3}$ power laws,~\cite{mcclain_gaussian-based_2017} placing them in the context of the general expectation that correlation energies, as a component of the total energy, would follow an $N^{-1}$ power law.~
\cite{fraser_finite-size_1996, williamson_elimination_1997, kent_finite-size_1999, lin_twist-averaged_2001, chiesa_finite-size_2006, gaudoin_quantum_2007, drummond_finite-size_2008, dornheim_ab_2016, gurtubay_benchmark_2010, ruggeri_correlation_2018, holzmann_theory_2016} %
We discussed but ultimately left open how this relates to correction terms which are popular elsewhere in the literature ~\cite{kent_finite-size_1999, gaudoin_quantum_2007, kwon_effects_1998, foulkes_quantum_2001, fraser_finite-size_1996, williamson_elimination_1997, holzmann_theory_2016, holzmann_finite-size_2011, drummond_finite-size_2008, chiesa_finite-size_2006, dornheim_ab_2016} and the $N^{-2/3}$ term in the exchange energy~\cite{drummond_finite-size_2008, kwee_finite-size_2008}.
We argue that our work is important even in the context of these approaches, as it shows that coupled cluster theory has the same power law dependence at the doubles level as these other total energy methods.

\section{Acknowledgements}

We gratefully acknowledge the University of Iowa for funding and computer resources through the University of Iowa Informatics Initiative. 
We are also grateful to Neil Drummond, Matthew Foulkes, Andreas Gr\"uneis, and Lubos Mitas for their emails and comments on this work. 
We gratefully acknowledge an email from Pablo Lopez Rios and Ali Alavi in response to our original arXiv preprint providing clarification around  Ref.~\onlinecite{ruggeri_correlation_2018}.
We thank David Ceperley and Richard Needs for useful conversations on related topics.
The data set used in this work is available at Iowa Research Online at URL: [to be inserted upon publication].

\providecommand{\latin}[1]{#1}
\makeatletter
\providecommand{\doi}
  {\begingroup\let\do\@makeother\dospecials
  \catcode`\{=1 \catcode`\}=2 \doi@aux}
\providecommand{\doi@aux}[1]{\endgroup\texttt{#1}}
\makeatother
\providecommand*\mcitethebibliography{\thebibliography}
\csname @ifundefined\endcsname{endmcitethebibliography}
  {\let\endmcitethebibliography\endthebibliography}{}

\end{document}